\documentclass[10pt,conference]{IEEEtran}

	\usepackage{ifpdf}

\usepackage{cite}

\ifCLASSINFOpdf
	\usepackage[pdftex,hiresbb]{graphicx}
\else
  \usepackage[dvips]{graphicx}
\fi

\usepackage[cmex10]{amsmath}
\usepackage{url}

\usepackage{enumitem}
\usepackage{multirow}
\usepackage{bigstrut}
\usepackage{amssymb,amsfonts}
\usepackage{bm}
	\usepackage[pdftex]{xcolor}
\usepackage[normalem]{ulem}
\usepackage{xspace}
\usepackage{textcomp}
\usepackage{algorithm}
\usepackage{algpseudocode}
\algrenewcommand\alglinenumber[1]{\footnotesize #1:}
\algrenewcommand\algorithmicindent{1.0em}
\usepackage{xpatch}
\usepackage[caption=false]{subfig}
\usepackage{placeins}
\usepackage{booktabs}

\usepackage{xurl}
\makeatletter
\newcommand{\removelatexerror}{\let\@latex@error\@gobble}
\makeatother

\DeclareMathOperator{\argmin}{argmin}

\makeatletter
\xpatchcmd{\algorithmic}{\itemsep\z@}{\itemsep=0.6ex plus 0.6ex  minus 0.6ex}{}{}
\makeatother

\usepackage{lettrine}
\newcommand{\introinitialdrop}[2]{%
  \lettrine[lines=2,lhang=0.0,nindent=0em]{#1}{\textsc{#2}}%
}

\makeatletter
\newcommand\fs@betterruled{%
  \def\@fs@cfont{\bfseries}\let\@fs@capt\floatc@ruled
  \def\@fs@pre{\vspace*{5pt}\hrule height.8pt depth0pt \kern2pt}%
  \def\@fs@post{\kern2pt\hrule\relax}%
  \def\@fs@mid{\kern2pt\hrule\kern2pt}%
  \let\@fs@iftopcapt\iftrue}
\floatstyle{betterruled}
\restylefloat{algorithm}
\makeatother

\newif\ifcleanversion
\cleanversiontrue

\ifcleanversion
  \AtBeginDocument{\colorlet{red}{black}}
  \renewcommand{\marginpar}[1]{}
\fi

\usepackage{booktabs}
\usepackage{tabularx}

\begin{document}
\title{Hamiltonian Monte Carlo for Vector Perturbation Precoding in MU-MIMO via Continuous Relaxation}

\author{
\IEEEauthorblockN{Junichiro Hagiwara\IEEEauthorrefmark{1},
Toshihiko Nishimura\IEEEauthorrefmark{2},
Yasutaka Ogawa\IEEEauthorrefmark{2}, and
Takeo Ohgane\IEEEauthorrefmark{2}}
\IEEEauthorblockA{\IEEEauthorrefmark{1}School of Social Informatics, Mukogawa Women's University,
Nishinomiya, Japan\\
Email: hagiwara\_junichiro\_x@mukogawa-u.ac.jp}
\IEEEauthorblockA{\IEEEauthorrefmark{2}{Faculty} of Information Science and Technology, Hokkaido University,
Sapporo, Japan\\
Email: \{nishim, ogawa, ohgane\}@ist.hokudai.ac.jp}
}

\maketitle

\begin{abstract}
Multi-user multiple-input multiple-output (MU-MIMO) is a key technology that improves wireless capacity through multiple antennas.
In MU-MIMO downlink precoding, vector perturbation (VP) is a representative nonlinear method that achieves high performance.
However, its search for the integer perturbation vector reduces to a closest vector problem, whose complexity grows rapidly as the number of users increases.
We propose a method that relaxes the discrete structure of the integer perturbation into a continuous mixture of $t$-distributions, enabling efficient search via gradient-based Hamiltonian Monte Carlo (HMC).
Complexity analysis and numerical experiments demonstrate the effectiveness of the proposed method.
Its search complexity scales as $\mathcal{O}(N^2)$ in the number of users $N$.
At a symbol error rate of $10^{-3}$, it performs within 2.4 dB of a hypersphere approximation benchmark, which approximates the performance limit of VP.
This paper reframes the VP perturbation search as a probabilistic inference problem, providing a general formulation for handling high-dimensional discrete search in a continuous space.
\end{abstract}

\begin{IEEEkeywords}
continuous relaxation, Hamiltonian Monte Carlo (HMC), multi-user multiple-input multiple-output (MU-MIMO), precoding, vector perturbation (VP).
\end{IEEEkeywords}

\section{Introduction} \label{sec:Introduction}
\introinitialdrop{T}{he} growing use of wireless communications, exemplified by 5G, calls for further improvements in spectral efficiency.
Multi-user multiple-input multiple-output (MU-MIMO)~\cite{MU-MIMO} is a promising multi-antenna technology that improves cell capacity without heavily relying on the implementation capabilities of user terminals.
In the MU-MIMO downlink, each user terminal cannot directly observe the channel states of other users.
Transmit precoding~\cite{Precoding}, which suppresses inter-user interference in advance at the base station, is therefore essential.
Precoding methods are broadly classified into linear and nonlinear methods.
Typical linear methods include zero forcing (ZF) and minimum mean squared error (MMSE)~\cite{chockalingam2014large}, while typical nonlinear methods include Tomlinson--Harashima precoding (THP)~\cite{THP-Tomlinson, THP-Harashima} and vector perturbation (VP)~\cite{VP-PartI, VP-PartII}.
In general, there is a trade-off: linear methods have low complexity but limited performance, whereas nonlinear methods achieve high performance at the cost of high complexity.
Although current deployments rely mainly on linear methods~\cite{3GPPTS38.214, Samsung2024mumimo}, developing nonlinear methods that achieve high performance with practical complexity is important for further capacity gains.
The theoretical performance limit of nonlinear precoding is given by dirty paper coding (DPC)~\cite{DPC}, and VP is recognized as a practical approximation of it.
In VP, an integer perturbation vector is added to the transmit symbols, and its effect is removed by a modulo operation at the receiver.
However, an exhaustive search for the integer perturbation vector requires a search cost that grows exponentially with the number of users, becoming infeasible in large-scale systems such as massive MIMO~\cite{An_Overview_of_Massive_MIMO}.

This perturbation search problem reduces to a closest vector problem, and various methods have been studied to solve it efficiently~\cite{Precoding}.
For example, sphere search~\cite{VP-PartII} achieves good performance by restricting the search candidates to lattice points within a hypersphere of a given radius, but it still incurs high computational complexity.
More recently, several methods have been proposed, including sequential perturbation updates based on a correlation metric~\cite{correlation_detection}, approximate message passing (AMP) applied to ZF and ZF with successive interference cancellation (SIC)~\cite{AMP_VP}, and quantum annealing that assumes dedicated hardware~\cite{Q_annealing1, Q_annealing2}.
These methods reportedly achieve good performance with high computational efficiency.
However, to the best of our knowledge, most existing methods treat the perturbation search essentially as a discrete optimization problem, and the direction of exploiting powerful algorithms developed for continuous spaces has not been sufficiently explored.

In this paper, we reframe the VP perturbation search as a probabilistic inference problem and develop a formulation for handling high-dimensional discrete search in a continuous space.
Specifically, we propose a method that relaxes the discrete structure of each integer perturbation element into a continuous mixture of $t$-distributions, enabling efficient search via gradient-based Hamiltonian Monte Carlo (HMC)~\cite{DUANE1987216}.
This method extends our preliminary study~\cite{RCS2023_VP}. 
In particular, we introduce lattice reduction (LR)~\cite{LR} as preprocessing.
LR improves the orthogonality of the lattice basis and thereby the conditioning of the perturbation space explored by HMC.
In related work, we have applied a similar probabilistic inference framework to MIMO signal detection at the receiver~\cite{Globecom}.
In contrast, this paper addresses a different problem, namely the VP perturbation search at the transmitter.
It provides a VP-specific formulation, including the construction of a likelihood based on the perturbation search criterion and the introduction of LR preprocessing.
Returning to our proposed method for VP, complexity analysis shows that its search scales as $\mathcal{O}(N^2)$ in the number of users $N$.
Numerical experiments show that, at a symbol error rate (SER) of $10^{-3}$, it performs within 2.4 dB of a hypersphere approximation benchmark~\cite{HA}.
This suggests that near-limit VP performance can be achieved in polynomial complexity on general-purpose computing platforms.
These results are expected to contribute to research and development toward 6G, as well as to broader insights into the continuous relaxation of combinatorial optimization.

The main contribution of this paper lies not in a new HMC algorithm, but in formulating the VP integer perturbation search as a differentiable posterior that is tractable with standard HMC.

The rest of this paper is organized as follows.
Section~\ref{sec:Problem formulation} formulates the problem, Section~\ref{sec:proposal} describes the proposed method, Section~\ref{sec:Numerical results and discussion} presents numerical results and discussion, and Section~\ref{sec:Conclusions} concludes the paper.

\textit{Notation:} $\boldsymbol{0}$ and $\boldsymbol{I}$ denote the zero vector and the identity matrix, respectively.
$\boldsymbol{A}^\mathrm{H}$ and $\boldsymbol{A}^\mathrm{T}$ denote the Hermitian transpose and the transpose of a matrix $\boldsymbol{A}$, respectively.
$\mathcal{N}(\boldsymbol{\mu}, \boldsymbol{\Sigma})$ denotes the normal distribution with mean vector $\boldsymbol{\mu}$ and covariance matrix $\boldsymbol{\Sigma}$.
$t_\nu(\mu, \sigma)$ denotes the $t$-distribution with location parameter $\mu$, scale parameter $\sigma$, and degrees of freedom $\nu$.
$\lVert \cdot \rVert$ denotes the Euclidean norm.
$Q(\cdot)$ denotes the nearest-integer quantizer.

\section{Problem formulation} \label{sec:Problem formulation}
\subsection{System Model}
We consider a MU-MIMO downlink system in which a base station with $M$ transmit antennas serves $N$ single-antenna user terminals.
The information symbol vector $\boldsymbol{u} \in \mathbb{C}^N$ has elements drawn independently and uniformly from the specified quadrature amplitude modulation (QAM) constellation.
Let $\tilde{\boldsymbol{u}} \in \mathbb{C}^M$ the precoded transmit symbol vector, $\boldsymbol{H} \in \mathbb{C}^{N \times M}$ the channel matrix, $\boldsymbol{y} \in \mathbb{C}^N$ the received symbol vector, and $\boldsymbol{w} \in \mathbb{C}^N$ the noise vector.
Their relationship is given by
\begin{align}
	\boldsymbol{y} = \boldsymbol{H} \tilde{\boldsymbol{u}} + \boldsymbol{w},
	\label{eq:mimosystem}
\end{align}
where $\boldsymbol{w}$ follows the complex normal distribution $\mathcal{CN}(\boldsymbol{0}, \allowbreak \sigma_w^2 \boldsymbol{I})$.
The total transmit power of the precoded vector is normalized to a given constant $P_t$, i.e., $\lVert \tilde{\boldsymbol{u}} \rVert^2 = P_t$.
In this paper, we assume that $\boldsymbol{H}$ is known at the transmitter.

\subsection{VP~\cite{VP-PartI, VP-PartII}}
\begin{figure}[!t]
\centering
\includegraphics[clip,width=1.0\linewidth]{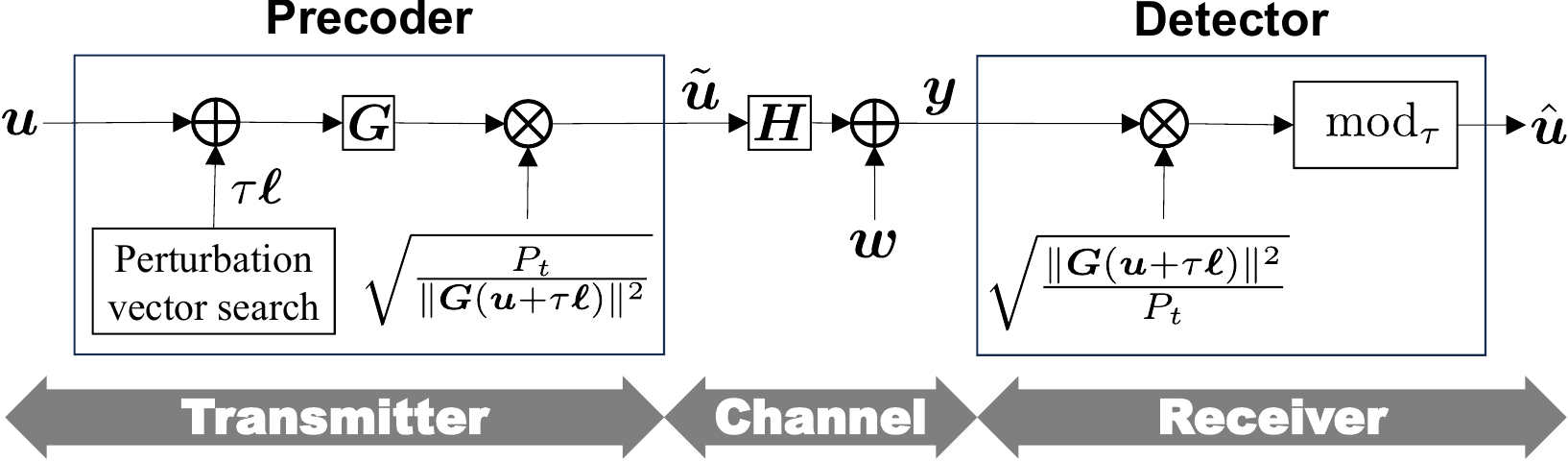}
\caption{Processing diagram of VP.}
\label{fig:VP}
\end{figure}

Fig.~\ref{fig:VP} shows the processing diagram of VP.
The precoder takes $\boldsymbol{u}$ as input and produces $\tilde{\boldsymbol{u}}$, while the detector takes $\boldsymbol{y}$ as input and produces $\hat{\boldsymbol{u}}$, the estimate of $\boldsymbol{u}$.
The precoder adds an integer perturbation vector to the information vector, and the detector removes the effect of the perturbation by a modulo operation.
The perturbation vector is chosen as a whole to improve the detection performance.
For symbols adversely affected by $\boldsymbol{H}$, the perturbation increases their transmit power to mitigate this effect.
Conversely, for symbols not adversely affected by $\boldsymbol{H}$, no perturbation is added.
Since the average transmit power is normalized to a constant, this effectively redistributes the transmit power to compensate for the adverse effect of $\boldsymbol{H}$.
$\tilde{\boldsymbol{u}}$ is given by
\begin{align}
	\tilde{\boldsymbol{u}} = \textstyle{\sqrt{\frac{P_t}{\lVert \boldsymbol{G} (\boldsymbol{u} + \tau \boldsymbol{\ell}) \rVert^2}}} \boldsymbol{G}(\boldsymbol{u} + \tau \boldsymbol{\ell}),
	\label{eq:encoder}
\end{align}
where $\boldsymbol{\ell}$ is the $N$-dimensional Gaussian integer perturbation vector, $\boldsymbol{G} = \boldsymbol{H}^\mathrm{H}(\boldsymbol{H}\boldsymbol{H}^\mathrm{H})^{-1}$ is the pseudo-inverse of $\boldsymbol{H}$, and $\tau = 2 c_\text{max} + \Delta$.
Here, $c_\text{max}$ is the absolute value of the constellation symbol(s) with largest magnitude, and $\Delta$ is the spacing between constellation points.

Similarly, $\hat{\boldsymbol{u}}$ is given by
\begin{align}
	\hat{\boldsymbol{u}} = \text{mod}_\tau(\boldsymbol{u} + \tau \boldsymbol{\ell} + \textstyle{\sqrt{\frac{\lVert \boldsymbol{G} (\boldsymbol{u} + \tau \boldsymbol{\ell}) \rVert^2}{P_t}}} \boldsymbol{w}),
	\label{eq:detector}
\end{align}
where $\text{mod}_\tau(\cdot)$ denotes the modulo operation with modulus $\tau$.
In this paper, for simplicity, we assume that $\lVert \boldsymbol{G} (\boldsymbol{u} + \tau \boldsymbol{\ell}) \rVert^2$ is known at the receiver.
From~\eqref{eq:detector}, the coefficient of $\boldsymbol{w}$ causes noise enhancement, and a smaller value leads to more accurate recovery of $\boldsymbol{u}$.
Therefore, the search criterion for $\boldsymbol{\ell}$ at the precoder is given by
\begin{align}
	\argmin_{\boldsymbol{\ell}} \lVert \boldsymbol{G} (\boldsymbol{u} + \tau \boldsymbol{\ell}) \rVert^2.
	\label{eq:VPcost}
\end{align}
Hereafter, $\lVert \boldsymbol{G} (\boldsymbol{u} + \tau \boldsymbol{\ell}) \rVert^2$ is referred to as the \textit{VP cost}.

Applying lattice reduction~\cite{LR} as preprocessing to improve the conditioning of $\boldsymbol{G}$, \eqref{eq:VPcost} becomes
\begin{align}
	  & \argmin_{\boldsymbol{\ell}} \lVert \boldsymbol{G} \boldsymbol{T} \boldsymbol{T}^{-1} (\boldsymbol{u} + \tau \boldsymbol{\ell}) \rVert^2 \notag \\
	= & \argmin_{\boldsymbol{\ell}_\text{LR}} \lVert \boldsymbol{G}_\text{LR} (\boldsymbol{u}_\text{LR} + \tau \boldsymbol{\ell}_\text{LR}) \rVert^2,
	\label{eq:VPcost_LR}
\end{align}
where $\boldsymbol{G}_\text{LR} = \boldsymbol{G} \boldsymbol{T}$, $\boldsymbol{u}_\text{LR} = \boldsymbol{T}^{-1} \boldsymbol{u}$, and $\boldsymbol{\ell}_\text{LR} = \boldsymbol{T}^{-1} \boldsymbol{\ell}$, and $\boldsymbol{T}$ is an $N \times N$ unimodular matrix, i.e., $\det(\boldsymbol{T}) = \pm 1$.
Lattice reduction transforms the basis of $\boldsymbol{G}$ into that of $\boldsymbol{G}_\text{LR}$, whose basis vectors are shorter and closer to orthogonal.
Therefore, searching for $\boldsymbol{\ell}_\text{LR}$ in \eqref{eq:VPcost_LR} and then transforming it back to $\boldsymbol{\ell}$ improves the search performance, compared with directly searching for $\boldsymbol{\ell}$ in \eqref{eq:VPcost}.
We adopt this approach in our method.
To obtain $\boldsymbol{T}$, we use the well-known LLL algorithm~\cite{LLL}.

\section{Proposed Method} \label{sec:proposal}
Minimizing the VP cost in \eqref{eq:VPcost_LR} is a high-dimensional combinatorial optimization problem, whose solution is not easily obtained.
To address this, we take an efficient approach that finds the solution through a probabilistic interpretation.
Finding $\boldsymbol{\ell}_\text{LR}$ for a given $\boldsymbol{u}_\text{LR}$ is probabilistically equivalent to inferring the posterior distribution $p(\boldsymbol{\ell}_\text{LR} \mid \boldsymbol{u}_\text{LR})$.
From Bayes' theorem, $p(\boldsymbol{\ell}_\text{LR} \mid \boldsymbol{u}_\text{LR}) \propto p(\boldsymbol{u}_\text{LR} \mid \boldsymbol{\ell}_\text{LR}) p(\boldsymbol{\ell}_\text{LR})$.
We will examine each component in turn. 
In this section, to simplify the representation of the perturbation elements, we use a real-valued representation equivalent to the complex one.
Here, the symbol notation remains unchanged, but the dimensions $N$ and $M$ in the complex representation are regarded as $2N$ and $2M$, respectively.

\subsection{Likelihood: $p(\boldsymbol{u}_\text{LR} \mid \boldsymbol{\ell}_\text{LR})$}
Since the VP cost has the form of a squared Euclidean norm, it can be probabilistically interpreted as a likelihood following a normal density, as follows:
\begin{align}
	p(\boldsymbol{u}_\text{LR} \mid \boldsymbol{\ell}_\text{LR}) = \mathcal{N}(\boldsymbol{u}_\text{LR}; -\tau \boldsymbol{\ell
}_\text{LR}, \sigma_\text{lik}^2(\boldsymbol{G}_\text{LR}^\mathrm{T} \boldsymbol{G}_\text{LR})^{-1}).
	\label{eq:lik}
\end{align}
Although $\sigma_\text{lik}$ equals $1$ by the definition of the VP cost, it affects the numerical search efficiency of the posterior distribution.
We therefore retain it as a tunable parameter and set an appropriate value in advance according to the problem.

\subsection{Prior Distribution: $p(\boldsymbol{\ell}_\text{LR})$}
It is known from~\cite{AMP_VP} that most effective $\boldsymbol{\ell}_\text{LR}$ concentrate around the reference point $Q(-\boldsymbol{u}_\text{LR}/\tau)$.
Considering the search efficiency, we restrict the deviation from the reference point to $\{-1, 0, 1\}$ in this paper.
The prior distribution therefore follows a three-peaked categorical distribution (Fig.~\ref{fig:pri} (a)).
We relax it into a three-peaked continuous mixture of $t$-distributions (Fig.~\ref{fig:pri} (b)), as follows:
\begin{figure}[!t]
\centering
\includegraphics[clip,width=0.9\linewidth]{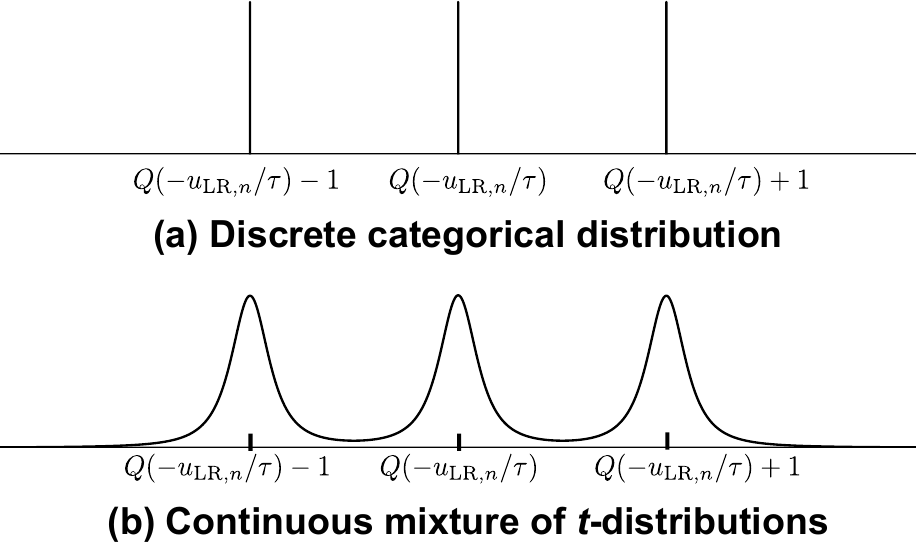}
\caption{Three-peaked categorical distribution and mixture of $t$-distributions.}
\label{fig:pri}
\end{figure}
\begin{align}
    p(\boldsymbol{\ell}_\text{LR}) &= \textstyle{\prod_{n=1}^{2N}} p(\ell_{\text{LR},n}), \notag \\
    p(\ell_{\text{LR},n}) &= \textstyle{\frac{1}{3} \sum_{i=-1}^{1}} t_\nu\!\left(\ell_{\text{LR},n};\, Q(-u_{\text{LR},n}/\tau) + i,\, \sigma_\text{pri}\right).
	\label{eq:pri}
\end{align}
$\nu$ and $\sigma_\text{pri}$ are tunable parameters, whose appropriate values must be determined in advance according to the problem.
We use a $t$-distribution rather than the typical normal distribution as the component distribution to improve the numerical search efficiency of the posterior distribution.
Since the $t$-distribution has a sharper mode and heavier tails than the normal distribution, it enables both intensive search around a mode and active movement across modes~\cite{Globecom}.

\subsection{Posterior Distribution: $p(\boldsymbol{\ell}_\text{LR} \mid \boldsymbol{u}_\text{LR})$}
Since both the likelihood and the prior distribution are continuous, the posterior distribution is also continuous.
Since the HMC described later operates on differences of the log-posterior distribution, we define the log-posterior distribution, retaining only the terms that depend on $\boldsymbol{\ell}_\text{LR}$, as follows:
\begin{align}
	\log p(\boldsymbol{\ell}_{\mathrm{LR}} \mid \boldsymbol{u}_{\mathrm{LR}}) & \stackrel{c}{=} 
		  - \frac{1}{\sigma_{\mathrm{lik}}^{2}} \left\| \boldsymbol{G}_{\mathrm{LR}} (\boldsymbol{u}_{\mathrm{LR}} + \tau\boldsymbol{\ell}_{\mathrm{LR}}) \right\|^{2}  \notag \\
		  & \quad +  \sum_{n=1}^{2N} \log \sum_{i=-1}^{1} a_{n,i}^{-(\nu+1)/2},		\label{eq:post}		\\
	a_{n,i} & =  1 + \frac{(\ell_{\mathrm{LR},n} - (Q(-u_{\text{LR},n}/\tau) + i))^{2}}{\nu\sigma_{\mathrm{pri}}^{2}}. \notag
\end{align}
Because the mixture prior is nonconjugate to the likelihood, an analytical point estimate of $\boldsymbol{\ell}_{\mathrm{LR}}$ is not available. We therefore explore the posterior numerically.
In this case, powerful algorithms developed for continuous spaces can be applied.
In this paper, we apply efficient HMC~\cite{DUANE1987216}, a type of Markov chain Monte Carlo (MCMC), to approximate the posterior distribution with a fixed number of samples.
The obtained samples are only candidates, and the perturbation search requires a point estimate.
Therefore, after quantizing them to the nearest integers, we finally select, as the final estimate, the one that minimizes the VP cost.

Algorithm~\ref{alg:hmc} outlines the HMC sampling procedure.
\begin{algorithm}[!tb]
\footnotesize
\caption{HMC sampling}
\label{alg:hmc}
\begin{algorithmic}[1]
\State Initialize $\boldsymbol{\ell}_\text{LR}$ at random
\For{$k = 1, \dots, L_\text{HMC}$}
	\State Draw $\boldsymbol{r}$ from $\mathcal{N}(\boldsymbol{0}, \boldsymbol{I})$
	\State \parbox[t]{1.0\linewidth}{Numerically solve Hamilton's equations to obtain $\boldsymbol{r}^\prime$ and $\boldsymbol{\ell}_\text{LR}^\prime$}
	\State \parbox[t]{0.95\linewidth}{Update $\boldsymbol{\ell}_\text{LR} \leftarrow \boldsymbol{\ell}_\text{LR}^\prime$ with probability min$\{1,  \exp(\mathcal{H}_{\mathrm{HMC}}(\boldsymbol{r}, \boldsymbol{\ell}_\text{LR}) \allowbreak - \mathcal{H}_{\mathrm{HMC}}(\boldsymbol{r}^\prime, \boldsymbol{\ell}_\text{LR}^\prime))\}$}
	\State \parbox[t]{0.9\linewidth}{Regard the updated $\boldsymbol{\ell}_\text{LR}$ as a sample from the posterior distribution $p(\boldsymbol{\ell}_\text{LR} \mid \boldsymbol{u}_\text{LR})$}
\EndFor
\end{algorithmic}
\end{algorithm}
Here, $L_\text{HMC}$ is the number of HMC iterations, and $\boldsymbol{r}$ is an artificially introduced auxiliary momentum variable that mitigates the risk of the $\boldsymbol{\ell}_\text{LR}$ search falling into a local optimum.
Letting $t$ denote a fictitious time, Hamilton's equations are ${\mathrm{d}\boldsymbol{\ell}_\text{LR}}/{\mathrm{d}t} = \boldsymbol{r}$ and $\mathrm{d}\boldsymbol{r}/{\mathrm{d}t} = \partial \log p(\boldsymbol{\ell}_\text{LR} \mid \boldsymbol{u}_\text{LR})/\partial \boldsymbol{\ell}_\text{LR}$.
This gradient information of the log-posterior distribution improves the efficiency of the $\boldsymbol{\ell}_\text{LR}$ search.
The Hamiltonian $\mathcal{H}_{\mathrm{HMC}}(\boldsymbol{r}, \boldsymbol{\ell}_\text{LR})$ is defined as the sum of the kinetic energy $\frac{1}{2}\lVert \boldsymbol{r} \rVert^2$ and the potential energy $-\log p(\boldsymbol{\ell}_\text{LR} \mid \boldsymbol{u}_\text{LR})$.
The HMC procedure itself, including the numerical integration and step-size/path-length settings, follows the standard implementation based on the Stan language~\cite{stan2023}.
Note that our contribution lies not in the HMC algorithm but in the differentiable posterior for the VP perturbation vector.

\section{Numerical results and discussion} \label{sec:Numerical results and discussion}
The purpose of this paper is to demonstrate the effectiveness of the proposed method based on the continuous relaxation approach.
To this end, we evaluate the SER through computer simulations and analyze the computational complexity.

\subsection{Common Assumptions}
In our simulations, the settings in Table~\ref{tbl:simulationassumption} are held constant across all SER evaluations.
\begin{table}[!tb]
\caption{Common simulation settings}
\label{tbl:simulationassumption}
\centering
{%
\footnotesize
\setlength{\tabcolsep}{3pt}
\renewcommand{\arraystretch}{1.05}
\begin{tabular}{@{}p{0.42\columnwidth}p{0.54\columnwidth}@{}}
\hline
Item & Setting \\
\hline
Number of SER evaluation trials
& $10{,}000$ for $N\in\{8,16\}$; $5{,}000$ for $N=32$ \\

Transmit antennas and users
& $M=N$, with $N\in\{8,16,32\}$ \\

Modulation
& 16QAM \\

Total transmit power
& $P_t=1$ \\

Fading model
& Uncorrelated quasi-static Rayleigh fading \\

Channel coding
& None \\
\hline
\end{tabular}
}%
\end{table}
Owing to space constraints, we report representative results for 16QAM; in preliminary experiments, we confirmed similar results with the same settings for other modulation orders such as 4QAM and 64QAM.

\subsection{Parameters of the Proposed Method (Table~\ref{tbl:proposed_param_settings})}
\begin{table}[!tb]
\caption{Parameter Settings for the Proposed Method}
\label{tbl:proposed_param_settings}
\centering
{%
\footnotesize
\setlength{\tabcolsep}{3pt}
\renewcommand{\arraystretch}{1.05}
\begin{tabular}{@{}p{0.64\columnwidth}p{0.28\columnwidth}@{}}
\hline
Parameter & Setting \\
\hline
Likelihood scale $\sigma_{\mathrm{lik}}$
& $0.72/\sqrt{N}$ \\

Prior scale $\sigma_{\mathrm{pri}}$
& $2.3/N$ \\

Degrees of freedom $\nu$
& $1.8$ \\

Number of parallel HMC chains
& $5$ \\

Number of HMC iterations $L_{\mathrm{HMC}}$
& $400$ \\
\hline
\end{tabular}
}%
\end{table}
The tunable parameters of the proposed method were determined by minimizing the SER in a preliminary validation.
When comparable SERs were obtained, we adopted the setting with lower complexity.
Since the VP search criterion does not depend on noise, the settings in Table~\ref{tbl:proposed_param_settings} were used commonly across all signal-to-noise ratio (SNR) points.

The likelihood and prior scale parameters $\sigma_\text{lik}$ and $\sigma_\text{pri}$ control how sharply the gradient of the log-posterior distribution is reflected in the HMC search, and thus strongly affect the search efficiency.
In this paper, we confirmed through preliminary validation that their optimal values exhibit a log-log linear trend with respect to the number of users $N$, and used the empirical formulas shown in Table~\ref{tbl:proposed_param_settings}.
The degrees of freedom of the mixture of $t$-distributions were searched over the grid $\{0.5, 0.6, \ldots, 4.9, 5.0\}$ and set to $1.8$.
To reduce the dependence on initial values and improve the stability of the search, we run multiple HMC chains in parallel.
The number of parallel chains was searched over the grid $\{1, 2, \ldots, 9, 10, 20, 50\}$ and set to $5$.
Finally, the number of HMC iterations $L_\text{HMC}$ was set to $400$, considering the balance between search performance and computational complexity.
We verify its validity in the following.

\begin{figure}[!t]
\centering
\includegraphics[clip,width=0.9\linewidth]{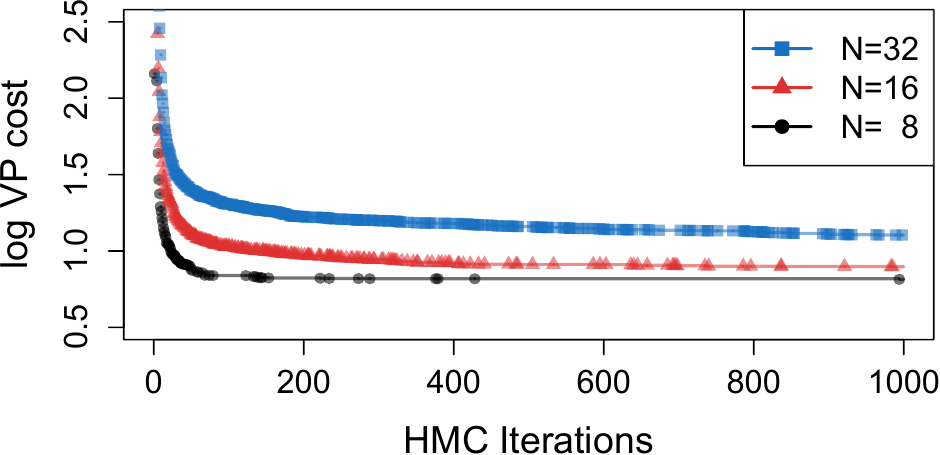}
\caption{Minimum log VP cost achieved up to each HMC iteration (median over 600 trials).}
\label{fig:L_HMC}
\end{figure}

A larger $L_\text{HMC}$ improves the search performance but increases the computational complexity, so their balance must be considered.
Fig.~\ref{fig:L_HMC} shows the evolution of the minimum log VP cost achieved up to each HMC iteration.
The values in the figure are the medians over 600 trials, obtained under the conditions in Tables~\ref{tbl:simulationassumption} and \ref{tbl:proposed_param_settings} (except for \textit{Number of SER evaluation trials} and $L_\text{HMC}$).
The saturation level increases with $N$ because the VP cost is an unnormalized quantity that includes the dimensional scale.
This does not affect the relative saturation assessment of the improvement, where the value at 1000 iterations is regarded as the convergence reference.
In Fig.~\ref{fig:L_HMC}, the additional improvement beyond 400 iterations was $0.3\%$, $1.3\%$, and $3.3\%$ of the total improvement up to 1000 iterations for $N = 8, 16, 32$, respectively (reflecting the greater search difficulty at larger $N$).
We therefore judged that the update of the minimum has largely saturated at 400 iterations, and stopped at 400 iterations as a sufficient value considering the computational complexity.
For completeness, convergence of the Markov chains was verified using standard diagnostics (e.g., the Gelman--Rubin statistic~\cite{BDA3}); details are omitted as they are not central to this paper.

\begin{figure*}[!tb]
\centering
	\begin{minipage}[t]{0.32\textwidth}
	  \centering
	  \includegraphics[clip,page=1,width=\linewidth]{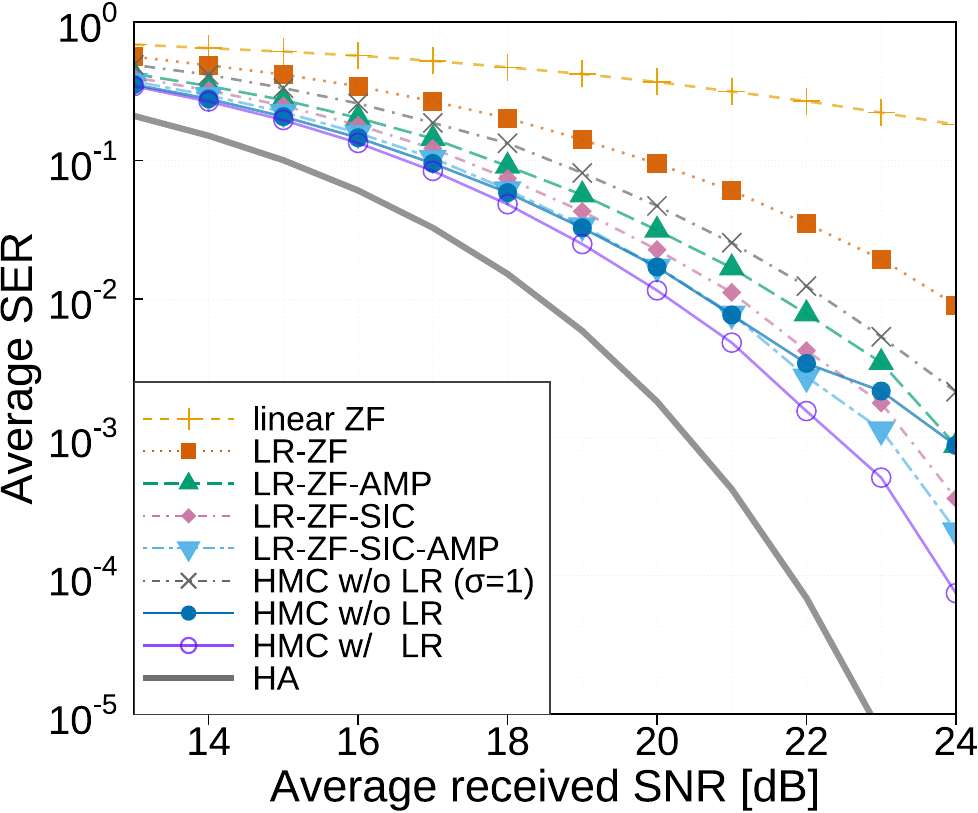}
	  \par\vspace{1mm}
	  {\footnotesize (a) $M=N=8$}
	\end{minipage}
	\hfill
	\begin{minipage}[t]{0.32\textwidth}
	  \centering
	  \includegraphics[clip,page=2,width=\linewidth]{./SER-crop.pdf}
	  \par\vspace{1mm}
	  {\footnotesize (b) $M=N=16$}
	\end{minipage}
	\hfill
	\begin{minipage}[t]{0.32\textwidth}
	  \centering
	  \includegraphics[clip,page=3,width=\linewidth]{./SER-crop.pdf}
	  \par\vspace{1mm}
	  {\footnotesize (c) $M=N=32$}
	\end{minipage}
\caption{Average SER versus average received SNR.}
\label{fig:SER}
\end{figure*}

\subsection{Comparison Methods}
As a representative basic method, we begin with linear ZF, which does not apply VP.
As representative statistical VP methods, based on~\cite{AMP_VP}, we consider four methods: LR-aided ZF, LR-aided ZF-SIC, LR-aided ZF with AMP, and LR-aided ZF-SIC with AMP, hereafter referred to as LR-ZF, LR-ZF-SIC, LR-ZF-AMP, and LR-ZF-SIC-AMP, respectively.
For AMP, we adopt the better-performing setting reported in~\cite{AMP_VP}.
As a performance benchmark for VP, we consider the characteristic based on the hypersphere approximation (HA)~\cite{HA}, which replaces the residual region of the closest vector problem solved by exhaustive-search VP with a hypersphere of the same volume.
This is hereafter referred to as HA.
Although HA is not a strict lower bound, it is known to accurately approximate the performance limit of VP~\cite{HA}, and we use it as a reference for the achievable performance in this paper.
Specifically, HA is evaluated by a large-system asymptotic expression with a fixed ratio of transmit to receive antennas, and is therefore shown as a constant SNR shift from the SISO AWGN (single-input single-output additive white Gaussian noise) characteristic.
Since the proposed method combines HMC with LR, it is hereafter referred to as HMC w/ LR.
For comparison, we also consider two variants without LR, obtained by setting $\boldsymbol{T} = \boldsymbol{I}$: one with the $\sigma$ settings in Table~\ref{tbl:proposed_param_settings}, and one with $\sigma_\text{lik} = \sigma_\text{pri} = 1$, unlike in Table~\ref{tbl:proposed_param_settings}.
These are hereafter referred to as HMC w/o LR and HMC w/o LR ($\sigma=1$), respectively.
Note that quantum annealing relies on dedicated hardware and different assumptions from this paper, so it is excluded from the comparison.

\subsection{SER Performance (Fig.~\ref{fig:SER})}
The performance of linear ZF degrades as $N$ increases.
This is because, as $N$ increases, small channel eigenvalues are more likely to appear, which in turn increases the risk of noise enhancement~\cite{Tse}.

Among the methods proposed in~\cite{AMP_VP}, the performance improves in the order LR-ZF, LR-ZF-AMP, LR-ZF-SIC, and LR-ZF-SIC-AMP.
For all methods, the performance degrades as $N$ increases, and AMP becomes less effective.
In particular, at $N=32$, the performances of LR-ZF-SIC and LR-ZF-SIC-AMP almost overlap.
This is presumably because small channel eigenvalues are more likely to appear as $N$ increases.
Specifically, the Gaussian assumption on the AMP residual is increasingly violated.

For the proposed method, we first examine HMC w/o LR ($\sigma=1$).
The degradation becomes more severe as $N$ increases, and the performance remains around that of LR-ZF.
This is because, as $N$ increases, the channel hardening effect~\cite{chockalingam2014large} reduces the relative channel-dependent fluctuation in the VP cost, so that a narrower search around the prior modes becomes sufficient.
As a result, the optimal values of $\sigma_\text{lik}$ and $\sigma_\text{pri}$ are considered to decrease as $N$ increases, as shown in Table~\ref{tbl:proposed_param_settings}, whereas fixing both to $1$ degrades the performance.
Next, we examine HMC w/o LR, with $\sigma_\text{lik}$ and $\sigma_\text{pri}$ set to their optimal values.
The performance improves substantially over HMC w/o LR ($\sigma=1$), but a floor tends to remain in the high-SNR region.
At high SNR, the AWGN amplitude becomes smaller, so the VP cost has a relatively larger effect on noise enhancement.
As a result, the search performance for the perturbation vector becomes more apparent.
Next, examining HMC w/ LR, we find that, except for a marginal reversal in the low-to-mid SNR region at $N=32$, it largely improves the performance of HMC w/o LR and also eliminates the floor in the high-SNR region.
This indicates that applying LR reduces the difficulty of the search and enables better perturbation vectors to be found.
Furthermore, comparing HMC w/ LR with LR-ZF-SIC-AMP, we find that HMC w/ LR outperforms it at all SNRs.
Specifically, the gain at $\text{SER} = 10^{-3}$ is $0.7$, $1.2$, and $1.0$ dB for $N = 8, 16, 32$, respectively.
We further compare HMC w/ LR with HA.
The degradation from HA at $\text{SER} = 10^{-3}$ is limited to $2.0$, $1.5$, and $2.4$ dB for $N = 8, 16, 32$, respectively.
Thus, HMC w/ LR shows good results, although its performance improves from $N=8$ to $16$ and degrades at $N = 32$.
This may be attributed to the competing effects of channel hardening and small eigenvalues, combined with the fixed $L_\text{HMC}$.
In particular, we preliminarily confirmed that the degradation at $N=32$ is partly related to the limited $L_\text{HMC}$, as suggested by the weaker saturation in Fig.~\ref{fig:L_HMC}; the full clarification of this $N$ dependence is left for future work.
In summary, we confirmed that, on general-purpose computing platforms, the proposed continuous relaxation approach outperforms conventional AMP-based methods and achieves an SER close to the performance limit of VP.

\subsection{Computational Complexity}
\begin{figure}[!tb]
\centering
  \begin{minipage}[t]{0.32\textwidth}
    \vspace{0pt}
    \centering
    \includegraphics[clip,page=1,width=\columnwidth]{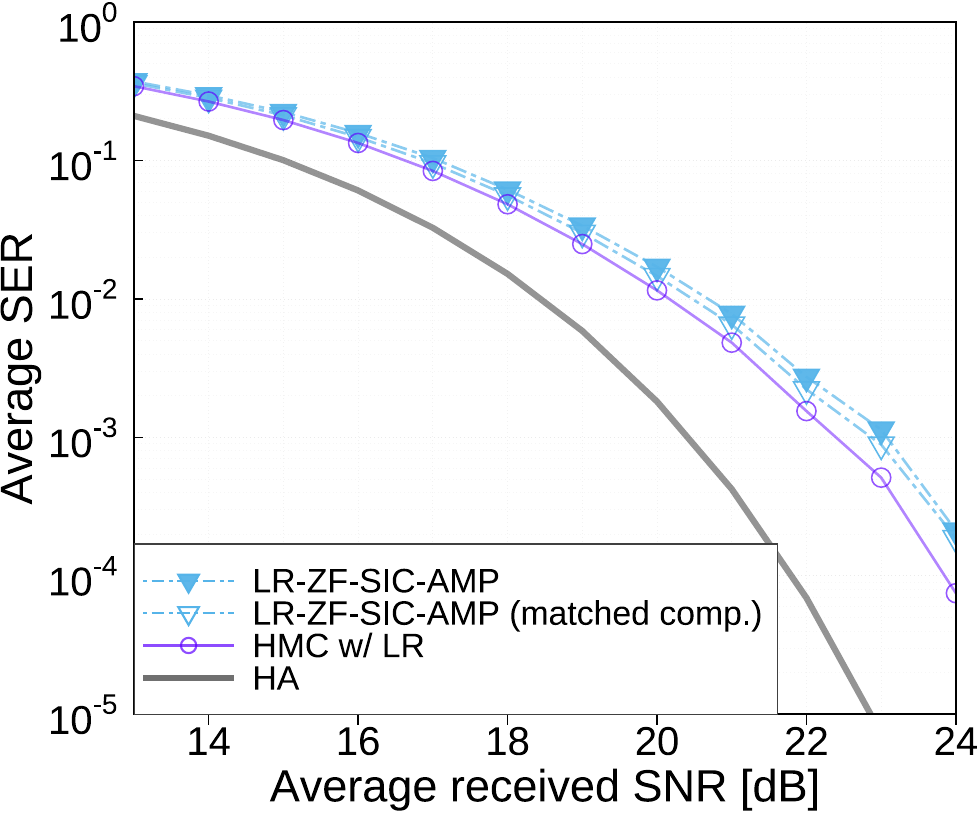}
  \end{minipage}
\caption{LR-ZF-SIC-AMP with matched complexity versus HMC w/ LR $(M=N=8)$.}
\label{fig:AMP33}
\end{figure}

We compare the number of multiplications in the iterative search between the proposed method and LR-ZF-SIC-AMP, excluding one-time preprocessing common to both.
In the proposed method, the dominant cost is the matrix-vector product $(\boldsymbol{G}_{\mathrm{LR}}^{\mathrm H}\boldsymbol{G}_{\mathrm{LR}}) \boldsymbol{\ell}_{\mathrm{LR}}$ in the evaluation of the gradient of the log-posterior distribution, which requires $N^2$ operations, since $(\boldsymbol{G}_{\mathrm{LR}}^{\mathrm H}\boldsymbol{G}_{\mathrm{LR}})$ can be computed once in advance.
This gradient evaluation is performed $L~(\ge 1)$ times when numerically solving Hamilton's equations in line 4 of Algorithm~\ref{alg:hmc}.
Since $L$ varies adaptively under the default settings of Stan, we use a representative value for the estimation.
In this paper, since the number of parallel HMC chains is $5$ and $L_\text{HMC} = 400$, the complexity is $5 \cdot 400 \cdot L N^2$.
In contrast, the dominant cost in LR-ZF-SIC-AMP is the matrix-vector product in the AMP processing, which requires $MN$ operations; since $M = N$ in our numerical experiments, this amounts to $N^2$ operations.
This operation appears at three places in the AMP processing loop~\cite[Alg. 1]{AMP_VP}: lines 6\&7, 8, and 10.
Based on~\cite{AMP_VP}, the number of AMP loops is set to $20$, so the complexity is $20 \cdot 3 \cdot N^2$.
Comparing the two methods, both are of polynomial order $\mathcal{O}(N^2)$, but the proposed method is computationally heavier than LR-ZF-SIC-AMP.
Here, we examine the performance of LR-ZF-SIC-AMP when the number of AMP loops is increased so that the complexity is approximately matched.
For a conservative comparison with LR-ZF-SIC-AMP, we consider $M=N=8$, where the AMP processing is most effective.
Since the median of $L$ in the simulation of Fig.~\ref{fig:SER}(a) was 15, we increase the number of AMP loops up to $5 \cdot 400 \cdot 15/3 = 10000$.
This performance comparison is shown in Fig.~\ref{fig:AMP33}.
The performance of LR-ZF-SIC-AMP is almost saturated and improves only marginally, so the proposed method still outperforms it.
This suggests that the performance gain of the proposed method is not merely due to the allocation of computational resources, but stems from the inherent effectiveness of the method.
That is, the proposed method extends the achievable SER performance.

\section{Conclusion} \label{sec:Conclusions}
This paper has shown that the proposed continuous relaxation approach can achieve performance close to the performance limit of VP in the MU-MIMO downlink, on general-purpose computing platforms and with polynomial complexity.
This result indicates a promising direction toward the practical use of high-performance nonlinear precoding for 6G wireless communications.
In this paper, the parameters affecting the performance of the proposed method were optimized empirically, and clarifying their mathematical background is left for future work.
This paper also recasts the integer perturbation search as a probabilistic problem in a continuous space, achieving good performance even in high-dimensional systems.
We have applied a similar probabilistic inference framework to MIMO signal detection at the receiver~\cite{Globecom}, demonstrating that this framework has the versatility to be applied to both transmitter- and receiver-side problems.
These results are expected to contribute to broadening the knowledge of continuous relaxation for combinatorial optimization.


\appendices




\end{document}